\documentstyle[aps,prd,floats,epsfig]{revtex}
\begin{document}

\title{Constraint-preserving boundary conditions in numerical relativity}

\author{Gioel Calabrese$^{1}$\thanks{gioel@gravity.phys.psu.edu}, Luis Lehner$^{2}$
\thanks{luisl@physics.ubc.ca} Manuel Tiglio$^{1,3}$\thanks{tiglio@gravity.phys.psu.edu}}

\address{1. Center for Gravitational Physics and Geometry, Department of Physics,\\ The
Pennsylvania State University, University Park, PA
16802.}
\address{2.
Department of Physics and Astronomy  \\
\& Pacific Institute for the Mathematical
Sciences\\
The University of British Columbia, Vancouver, BC V6T 1Z1}
\address{3. Center for Gravitational
Wave Physics, Department of Physics,\\ The
Pennsylvania State University, University Park, PA
16802.}

\maketitle

\begin{abstract}
This is the first paper in a series aimed to implement boundary 
conditions consistent with the constraints' propagation in 3D unconstrained
numerical relativity. Here we consider spherically symmetric black hole
spacetimes in vacuum or with a minimally coupled scalar field, within the
Einstein-Christoffel (EC) symmetric hyperbolic formulation of Einstein's equations. By
exploiting the characteristic propagation of the main variables and constraints,
we are able to single out the only free modes at the outer boundary for these problems.
In the vacuum case a single free mode exists which 
corresponds to a gauge freedom, while in the matter case an extra mode exists which is
associated with the scalar field. We make use of the fact that the EC formulation has no
superluminal characteristic speeds to excise the singularity.

We present a second-order, finite difference discretization to treat
these scenarios,
where we implement these constraint-preserving boundary conditions, and are able to evolve
the system for essentially unlimited times. As a test of the robustness of our
approach, we allow large pulses of gauge and scalar field to enter the domain through the
outer boundary. We reproduce expected results, such as trivial (in the physical sense)
evolution in the vacuum case (even in gauge-dynamical simulations), and the tail decay for
the scalar field. 
\end{abstract}

\section{Introduction and overview}

In the initial value problem of Einstein's equations one decomposes the original
system of equations (given by $G_{ab}=8\pi T_{ab}$) into two distinct sets: one set consisting
of {\it evolution equations} (involving time derivatives of the main variables) and the
other consisting of {\it constraint equations} (which do not involve time derivatives).
 There are infinite possible ways of achieving this decomposition, and the resulting
systems are known by a variety of different names (e.g. ADM, 
characteristic-Bondi and conformal-Einstein approaches, to name just three).
However, depending on the character of the hypersurfaces used to foliate the
spacetime, these formulations can be seen as belonging to different groups: the
group of Cauchy formulations (which require a spacelike foliation), that of
characteristic formulations (having a null foliation), or a more generic group
(where the foliation's leaves need not have a fixed specific character) (see, for
instance,~\cite{luisreview}).

Irrespective of the group and restricting to the initial value problem (where the
problem is boundary-free), the state of the system is defined at an initial
hypersurface $\Sigma_0$ and the solution to the future of $\Sigma_0$ is obtained through
 the evolution equations. However, if the system of equations involves
constraints, one might opt to employ only a subset of the evolution equations supplemented
with enough constraint equations to solve for the field variables. This strategy is
referred to as {\it constrained evolution}, as opposed to {\it free evolution} (where only
the evolution equations are employed). In the particular case of Einstein's equations, a
straightforward manipulation of the Bianchi identities demonstrates that either strategy
produces, at the analytical level, the same solution. Therefore, one need not deal with
constrained evolution (which usually requires solving elliptic equations) and the more
direct approach of free evolution can be safely employed. In the numerical realm, however,
the picture is more complicated (for definiteness, from now on we concentrate on the case
of Cauchy evolution). On one hand, employing constrained evolutions might represent a
significant computational overhead as it usually involves solving elliptic equations at
each time step; for this reason constrained evolutions have been, for
the most part, avoided beyond the two-dimensional case. On the other hand, free
evolution in numerical
implementations (which only evaluate the constraint equations to monitor the quality of
the implementation) display violation of the constraints. At early times, these violations
are consistent with the truncation error~\cite{choptuikconsistency}, but as
time progresses the observed violations often grow quite rapidly.

The picture gets even more complicated in the presence of boundaries. The violation of
the constraints in unconstrained numerical evolutions frequently grows without bound.
Possible reasons, besides the ones already present in boundary-free models, are
constraint-violating modes introduced by standard boundary conditions (which
might drive instabilities, and in any case do introduce spurious errors that should be
avoided). If the case under study corresponds to cosmological scenarios or short term
evolutions, the aforementioned problem should not be worrisome. In the cosmological case
one has no boundaries (or rather, periodic boundary conditions are specified, which
trivialize the specification issue), for solutions which are restricted to the future
domain of dependence of the initial data no boundary conditions need be specified, and for
short-term evolutions the errors introduced by approximate boundary conditions
 are restricted to a small region. There are, however, many
non-cosmological important problems that require long term evolutions and hence the need
to prescribe consistent boundary conditions; among these, collapse scenarios and black
hole spacetimes.

In the one-dimensional case, approximate boundary conditions that minimize
the influence of errors introduced at the boundary have
been presented~\cite{1d_bound}. This is achieved by placing the boundaries
far away and treating a region close to them in a special way (to control spurious
reflections). Unfortunately, these techniques are not as effective in higher dimensional
scenarios, as the computational cost involved in placing boundaries far away is excessive.
Furthermore, even when there were enough computational resources, it would be preferable to
make use of them to achieve finer resolved simulations, rather than to push boundaries
farther away. It is therefore of considerable importance to formulate relatively
inexpensive boundary conditions whose associated errors do not depend on the boundaries'
locations, minimize spurious reflections, and guarantee that constraint-violating modes
are not introduced.

Essentially, one aims to provide boundary conditions such that there is not only a unique
solution to the evolution equations, but there is {\em also} preservation of the
constraints throughout the computational domain.
In a strongly hyperbolic formulation of
gravity (see \cite{hyp_reviews} for reviews on hyperbolic techniques applied to Einstein's
equations) the first requirement amounts to giving boundary conditions only to
the characteristic modes that enter the computational domain at any given boundary.
Satisfying the second requirement is considerably more involved and has only very recently
started receiving attention. In the analytical realm, well posedness of the initial
boundary value problem (for a particular system) has been established
in~\cite{friedrichnagy}, which sheds light on the physical understanding of the issue and
shows that, at least in a by-case basis, the problem might be analytically tractable. In
the numerical realm, several efforts (restricted to different scenarios) have illustrated
the advantages of providing boundary conditions through the use of constraints~\cite{ibv}.
It is precisely this problem  that we want to address here within the context of Cauchy,
unconstrained evolution.

The present work aims to contribute to the area with the particular motivation of
addressing issues proper of numerical implementations. Although in this first paper we
restrict our analysis to the spherically symmetric case and a particular way of
reformulating Einstein's equations, the procedure is straightforward to
implement within any hyperbolic formulation of gravity and in three dimensions
(3D).

We present a detailed discussion of the analytic treatment for the boundary conditions and
illustrate its benefits with a simple numerical implementation which accurately evolves a
spherically symmetric spacetime, stationary or dynamical (the dynamics either
corresponding to gauge modes in the vacuum case or to true dynamics in the case of a
minimally coupled scalar field), for unlimited periods of time and
without sign of instabilities. Throughout this work we minimize the use of
techniques specifically developed for treating spherically symmetric spacetimes, both
 analytical and numerical, to expedite the generalization to the 3D
case.

Our starting point is the assumption that we are dealing with a first-order, quasilinear
strongly  hyperbolic formulation for Einstein's equations\footnote{In  3D, a generic
first-order   system can be written as $\dot{u} = \sum_{i=1}^3 A^i \partial_i u +
l.o.$; defining  the symbol $P:=A^i \omega _i$, such a system is said to be
{\it symmetric} (or symmetrizable, the terminology in the literature is not uniform)
{\it hyperbolic} if $P_{|\omega =1|}$ can be symmetrized, by a smooth transformation
independent of $\vec{\omega} := (\omega _1, \omega _2, \omega _3)$, while it is said to be
{\it strongly hyperbolic} if it can be symmetrized (the symmetrizer is not required to be
independent of $\vec{\omega} $), and its eigenvalues are real. In spherical symmetry, a
strongly hyperbolic system is symmetrizable (the converse statement is always true), but
 this is an artifact of one dimension (1D).}. In the spherically symmetry case such a
system can be written as $\dot{u} = A u' + l.o.$, where $u$ is an array of variables, dot
and prime indicate time and spatial derivatives, respectively, $A$ (called the {\it
principal part}) is a diagonalizable matrix that might depend on $u$ and the spacetime
coordinates, but not on derivatives of $u$, and $l.o.$ stands for {\it lower order
terms}, i.e.~terms that do not have derivatives (of any kind). So, we have a system of
equations for the main variables\footnote{The word {\it main} will be used, when there is
possibility of confusion, to differentiate between the evolution equations for say,
the three-metric and extrinsic curvature (and possibly extra variables) and evolution for
the constraints.}, \begin{equation}
\dot {u} = Au' + l.o.   \;\;\; \mbox{(System I)} \label{system1}    \; ,
\end{equation}
and evolution for the constraints, which we assume are also
strongly hyperbolic,
$$
\dot{u}_c = A_c u_c' + l.o.  \;\;\; \mbox{(System II)} \; .
$$

If one were interested only in system I, one would give initial data for the variables
that form $u$, and boundary data only to the ingoing characteristic
modes\footnote{By {\it ingoing} we refer to those modes entering the computational domain
with respect to a given boundary.}. These
characteristic modes (eigenvectors of $A$) are, depending on which boundary one is dealing
with, the ones that are travelling to the left (positive eigenvalues) or right (negative
eigenvalues).

A unique solution to system II is fixed, similarly, by giving initial data to $u_c$ and
boundary conditions to the ingoing modes of the system. Since this is supposed to be an
homogeneous system (which is the case in Einstein's equations
\cite{frittelliCONSTRAINT}), the identically zero solution is obtained by providing zero
as initial data ($u_c=0$) and zero boundary conditions to the eigenmodes of $A_c$
entering the domain. The crucial point here is that, in unconstrained evolution, initial
data and boundary conditions for $u_c$ follow from that of $u$. Thus, one has to provide
initial and boundary data for $u$ such that they imply zero initial data and boundary
conditions for $u_c$ (i.e.~the data are consistent with all Einstein's equations). The
first step, namely, providing initial data that satisfy the constraints, is
customarily well satisfied as the constraint equations are employed in the determination
of data for $u$ (see~\cite{cookID,luisreview} and references cited therein). It is the
second step that is usually neglected. Here we concentrate on
this problem with the goal of providing consistent data, whose associated error neither
depends on the location of the boundaries nor on the total evolution length; rather its
error will agree with the overall truncation error of the global implementation.

The main difficulty with this second step is that the constraints (and, therefore the
eigenmodes of system II as well) involve spatial derivatives of the main variables.
On the other hand, by controlling the (ingoing) characteristic modes of system I at the
boundaries, one does not control the spatial derivatives needed to set the ingoing
constraint modes to zero. One way
around this (and the one we use here), is to trade spatial derivatives for time
derivatives using system I. Consequently, enforcing the constraints at the boundary
amounts to controlling some time derivatives. We will make this
construction explicit later but, before proceeding further, some comments are appropriate
(both issues certainly deserve further analysis):
\begin{itemize}
\item To our knowledge, there is no rigorous result guaranteeing
that in any strongly hyperbolic formulation of Einstein's equations this trading of
spatial for time derivatives can be done.

\item When performing the trading, one ends up with some conditions
at the boundaries on some time derivatives of the main variables. One must find
out whether these conditions can be fulfilled by controlling only the ingoing modes of
system I (at the numerical level, conditions on the time derivatives are enough
for the application of the method of lines~\cite{kreiss}). Again, we know of no proof
showing that this should be possible in a general case.
\end{itemize}
We here show how this inversion can be performed
in the 1D case, and leave for a future paper a similar analysis in 3D linear gravity
\cite{3dlinear}.

When dealing with black hole spacetimes, boundary conditions customarily refer to the
{\it outer boundary} conditions, but one might also have inner boundaries.
These 
constitute ``holes'' or ``excised'' regions from a given computational domain; the most
widely considered are those where the excised region has been chosen so as to remove
the singularities when dealing with black hole spacetimes (assuming the validity of cosmic
censorship). In this case, a region inside the black hole is excised from the
computational domain (Unruh, cited in\cite{unruh}). This excision strategy introduces an
inner boundary which, if chosen inside the black hole, should leave the region outside the
event horizon unaffected. Mathematically, the realization of this idea is ensured
 by employing a hyperbolic formulation with no superluminal characteristic speeds and, in
particular, where all eigenvalues describe modes propagating towards the inner
boundary\footnote{For a discussion of different approaches towards application of excision
techniques see~\cite{luisreview} and references cited therein.}. The non-triviality of
excision if one does {\em not} have a formulation with these properties is discussed in
one of the appendixes.

In recent years, a number of re-formulations of Einstein's equations with
physical characteristic speeds have been
presented~\cite{ay,fr,kidder2}: Here we make use of one of them in our analysis, the
so-called Einstein-Christoffel (EC) system~\cite{ay} (note that any other would have been
equally well suited - our choice is simply motivated by a comparison with available
results \cite{kidder1}).

The organization of this paper is the following: in section II we present the
equations for the EC system describing a spherically symmetric massless scalar field
minimally coupled to gravity. In section III we describe how to give boundary
conditions that preserve the constraints, while in section IV  we present our numerical
method to test the procedure. Numerical results are included in section V, and are
divided into three classes:
evolution of stationary slicings of a Schwarzschild black hole, evolution of
dynamical slicings of the same spacetime, and
evolution of a massless scalar field interacting with a black hole. In all cases the
simulations can be followed for unlimited times, even using very moderate resolutions. We demonstrate
the accuracy of our method by monitoring the mass of the black hole in the vacuum
case and reproducing the known tail decay in the scalar field case. A particularly strong
test of the robustness of the approach is presented by letting strong gauge or scalar field
pulses enter the computational domain through the outer boundary (compare with
standard outer boundary treatments, where conditions are obtained by requiring the geometry at the outer
boundary to be close to flat or Schwarzschild spacetimes, i.e. quite the opposite from
what we do here).

\section{Main evolution equations and propagation of the constraints}

As often is the case in hyperbolic formulations of Einstein's equations, the EC
formulation uses ``exact'' (i.e.~arbitrary but {\em a
priori} specified) shift and exact densitized lapse (defined by $\alpha = Ng^{-1/2}$,
where $g$ is the determinant of the three-metric and $N$ is the lapse). In the
present work, and for the sake of maintaining a uniform notation, we
will follow as much as possible the conventions of \cite{kidder1}. For example, we write
$\tilde{\alpha }:= \alpha r^2 \sin {\theta }$, and
\begin{eqnarray*}
ds^2 & = & -N^2dt^2 + g_{rr}(dr+\beta
dt)^2 + r^2 g_T(d\theta ^2 + \sin ^2{\theta } d\phi^2) \;,  \\
K_{ij}  & = &   K_{rr}dr^2 + r^2 K_T(d\theta ^2 + \sin ^2{\theta }
d\phi^2)                                                       \; ,
\end{eqnarray*}
where all fields depend only on $(t,r)$. Since the EC formulation is a first-order
reformulation of Einstein's equations, besides the three metric and extrinsic curvature,
further variables (which basically contain information of spatial derivatives of the
three-metric) are needed. In our present case of spherical symmetry, only two new
variables require introduction, $f_{rrr}$ and $f_{rT}$, defined by
\begin{eqnarray*}
f_{rrr} &=& \frac{g_{rr}'}{2}  + \frac{4g_{rr}f_{rT}}{g_T} \; ,  \\
f_{rT} &=& \frac{g_T'}{2} + \frac{g_T}{r}  \; . \\
\end{eqnarray*}
Additionally, we plan to study a scalar field $\Psi$ minimally coupled to the geometry. In
order to re-express the equation that governs this field,
$$
g^{ab}\nabla _a \nabla _b \Psi = 0 \; ,
$$
as a first-order hyperbolic system, we introduce two new variables:
\begin{eqnarray*}
\Pi & :=  & \frac{1}{\alpha }(\beta \Psi ' - \dot{\Psi}) \; ,\\
\Phi & :=  & \Psi ' \; .
\end{eqnarray*}
The evolution equation for $\Psi $ decouples from the rest, in the sense that one has a
closed system for the set of eight variables $(g_{rr}, g_T, K_{rr},K_T, f_{rrr},
f_{rT},\Pi,\Phi)$ which one can solve for, and afterwards obtain $\Psi$. Because
of this, and following common practice, we will drop $\Psi$ from the system.

Thus, the main evolution equations, up to principal part (the complete
expressions are listed in the appendix), are
\begin{eqnarray}
&& \dot{g}_{rr} = \beta g'_{rr} + l.o. \; , \label{grr_dot}\\
&& \dot{g}_T = \beta g'_T + l.o. \; , \label{gT_dot} \\
&& \dot{K}_{rr} = \beta K'_{rr} - \frac{N}{g_{rr}}
f'_{rrr} + l.o. \; , \label{Krr_dot} \\
&& \dot{K}_{T} = \beta K'_{T} - \frac{N}{g_{rr}} f'_{rT}
+ l.o., \label{KT_dot} \; , \\
&& \dot{f}_{rrr} = \beta f'_{rrr} - N K'_{rr} +  l.o. \; ,
\label{frrr_dot}\\
&& \dot{f}_{rT} = \beta f'_{rT} - N K'_T + l.o. \; ,
\label{frT_dot} \\
&& \dot{\Pi } = \beta \Pi ' - \frac{N}{g_{rr}} \Phi ' +  l.o. \; ,  \label{pidot}\\
&& \dot{\Phi } =  \beta \Phi ' - N\Pi ' + l.o. \;. \label{phidot}
\end{eqnarray}

The characteristic modes and eigenvalues determined
by the system play a crucial role in our boundary treatment. These are (note that the
modes with speed $\beta$ propagate along the timelike normal to the foliation, while the
other modes propagate along the light cone),
\begin{eqnarray}
& & u_1 = g_{rr} \;\;\; (v_1 = \beta) \; , \label{u1}\\
& &  u_2 = g_T \;\;\; (v_2 = \beta)\; , \label{u2} \\
& &  u_3 = K_{rr} - g_{rr}^{-1/2}f_{rrr} \;\;\; (v_3 = \beta +
\tilde{\alpha} 	g_T) \; , \label{u3} \\
& &  u_4 = K_{T} - g_{rr}^{-1/2}f_{rT} \;\;\;
(v_4 = \beta + \tilde{\alpha} g_T) \; , \label{u4} \\
& &  u_5 = K_{rr} + g_{rr}^{-1/2} f_{rrr} \;\;\; (v_5 = \beta -
\tilde{\alpha} g_T) \; , \label{u5} \\
& &  u_6 = K_{T} + g_{rr}^{-1/2} f_{rT} \;\;\;  (v_6 = \beta -
\tilde{\alpha} g_T) \; ,    \label{u6}           \\
& &  u_7 = \Pi + g_{rr}^{-1/2}\Phi \;\;\; (v_7 = \beta - \tilde{\alpha} g_T)\; , \\
\label{u7}
& &  u_8 =  \Pi - g_{rr}^{-1/2}\Phi \;\;\; (v_8 = \beta + \tilde{\alpha} g_T) \; .
\label{u8} \end{eqnarray}
If we were
dealing with a constraint-free system described by
equations (\ref{grr_dot}-\ref{phidot}), a unique solution would be fixed by
providing boundary conditions for the characteristic modes that are incoming at the
boundaries and data on the initial hypersurface.
However, we know that a solution to equations (\ref{grr_dot}-\ref{phidot}) is a solution
of Einstein's equations if and only if the following four constraints are also satisfied:

\begin{eqnarray}
& & C :=  \frac{f_{rT}'}{g_{rr}g_T}
- \frac{1}{2r^2g_T} + \frac{f_{rT}}{g_{rr}g_T} \left(
\frac{2}{r} + \frac{7 f_{rT}}{2 g_T} - \frac{f_{rrr}}{g_{rr}} \right) -
\frac{K_T}{g_T} \left(\frac{K_{rr}}{g_{rr}} + \frac{K_T}{2g_T} \right)  +
\frac{\Phi ^2}{4g_{rr}} + \frac{\Pi ^2}{4} = 0 ,
\label{C} \\
& & C_r := \frac{K_T'}{g_T} + \frac{2K_T}{rg_T} -
\frac{f_{rT}}{g_T} \left( \frac{K_{rr}}{g_{rr}} + \frac{K_T}{g_T} \right)
+ \Phi \Pi =0 , \label{Cr}\\
& & C_{rrr} := g_{rr}' + \frac{8g_{rr}f_{rT}}{g_T}
- 2f_{rrr} = 0, \label{Crrr} \\
& & C_{rT} := g_T' + \frac{2g_T}{r} -2f_{rT} = 0 \; . \label{CrT}
\end{eqnarray}
The first two are basically the Hamiltonian and momentum constraints, respectively, while
the other two correspond to the definitions of the extra variables that make the
system first order with respect to spatial derivatives.
As mentioned, it is common practice to choose consistent initial data for Einstein's
equations by solving these constraints; however the constraints have been
examined in a limited number of cases to provide consistent boundary
data. The main purpose of this work is to provide further
indications that constraints should be looked at more closely when dealing with boundary
conditions. First, note that these constraints are defined in terms of the main variables,
and a solution of the main evolution equations completely determines them as functions of
spacetime. In particular, one can obtain the time evolution of the constraints by: (i) taking time
derivatives of the right hand side (r.h.s.) of equations (\ref{C}-\ref{CrT}); (ii) replacing
the time derivatives by the r.h.s. of equations (\ref{grr_dot}-\ref{phidot});
and, (iii) re-expressing the main variables in terms of the constraints and their spatial
derivatives. In the present case, these are

\begin{eqnarray*}
& & \dot{C} = \beta C' - \frac{N}{g_{rr}}
C_r ' + l.o. \;,  \\
& & \dot{C}_r = - N C' + \beta C'_r + l.o. \;, \\
& & \dot{C}_{rrr} = \beta C'_{rrr} + l.o. \;,\\
& & \dot{C}_{rT} = \beta C'_{rT} + l.o. \; .
\end{eqnarray*}
We can also calculate the characteristic modes and
eigenvalues of this system, obtaining:
\begin{eqnarray}
& & C_1 = C + g_{rr}^{-1/2} C_r \;\;\;   (v^c_1 =
\beta - \tilde{\alpha} g_T), \label{c1}\\
& & C_2 = C - g_{rr}^{-1/2} C_r \;\;\;
(v^c_2 = \beta + \tilde{\alpha} g_T), \label{c2} \\
& & C_3 = C_{rrr} \;\;\; (v^c_3 =
\beta),\label{c3} \\
& & C_4 = C_{rT} \;\;\; (v^c_4 = \beta). \label{c4}
\end{eqnarray}
In the next section we exploit this information about the
characteristic structure of our system in order to set up boundary conditions
that will preserve the constraints at the boundaries.

\section{Constraint-preserving boundary conditions}

In any system where the characteristic modes and eigenvalues are known, the
data  required at a given boundary are straightforwardly assessed by examining the
eigenvalue of each characteristic mode. In our present case,
these eigenvalues are $\lambda _1= \beta - \tilde{\alpha}
g_T$, $\lambda _2 = \beta + \tilde{\alpha} g_T$, and $\lambda _3 = \beta $. In the case
 where the shift is exact, one can a priori guarantee which sign $\lambda _3$
will have. Throughout this paper we shall take $\beta$ as positive, since in that way we
will be able to reproduce stationary known slicings of the Schwarzschild
space-time\footnote{Kerr-Schild, Painlev\'e-Gullstrand, full harmonic and time harmonic
slicings of Schwarzschild have positive shifts, see ~\cite{kidder2} for the
explicit form of these metrics in the EC formulation.}, and to evolve dynamical spacetimes
as well. The positivity of the shift implies that $\lambda _2$ is also positive. On the
other hand, the sign of $\lambda _1$ depends on the solution and, thus, cannot be
controlled  {\it a priori}. However, for typical stationary slicings of Schwarzschild
it is negative (positive) outside (inside) the black hole, and zero at the horizon. By
continuity, the same will hold for (perhaps slight) distortions of these slicings. In our
simulations we check numerically that this is indeed the case. As we shall see, even on
highly distorted spacetimes this condition remains satisfied. Figure 1 shows a schematic
diagram for the characteristic modes of a Schwarzschild black hole.

\begin{figure}[ht]
\begin{center}
\epsfig{file=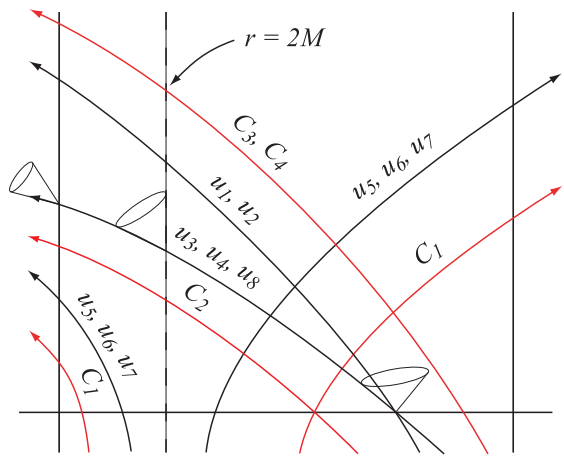,clip=,height=60mm}
\caption{Schematic diagram for the characteristic speeds of the different modes in a
stationary slicing of Schwarzschild black hole.}
\end{center}
\end{figure}

Our inner boundary is always inside the black hole, and no conditions
are needed there. The usual procedure to make sure that the inner boundary is
inside the hole is to track the apparent horizon. In spherical
symmetry this is particularly simple since its location is
defined in terms of {\em algebraic} combinations of the main variables ($ f_{rT} -
g_{rr}^{1/2}K_T = 0 $). Our actual procedure is to choose the initial data such that the inner
boundary is inside the apparent horizon, and then numerically monitor its position during
evolution. In our simulations the horizon never touches the inner boundary, so we do not
need to move it, i.e.~its grid location is fixed.

From the discussion in section II, we know that
 at the outer boundary (which in our case also has fixed grid location) we
  need to specify $u_1, u_2, u_3, u_4,u_8$ from the characteristic structure of
the main equations. However, since the system is constrained, we are not free to choose
these arbitrarily (which would be the case in an unconstrained system), but rather we must
do so such that $C_2=C_3=C_4=0$ are satisfied (Note that $C_1$ is outgoing, therefore
nothing special needs be done for it, nor should it be, otherwise we would overdetermine
the constraint system of evolution equations.). This procedure, coupled with initial data
satisfying the constraints, and the fact that the constraints are propagated with a
quasilinear homogeneous system, will ensure that they are preserved everywhere.

We now make explicit how this procedure is implemented in our present case. We start by
discussing how to enforce $C_3=C_4=0$ at the (outer, from now on) boundary. Writing down
the constraints explicitly in equations (\ref{c3},\ref{c4}), these conditions are
\begin{eqnarray}
g_{rr}' + \frac{8g_{rr}f_{rT}}{g_T} - 2f_{rrr} = 0, \\
g_T' + \frac{2g_T}{r} -2f_{rT} = 0 \, .
\end{eqnarray}
Using equations (\ref{grr_dot},\ref{gT_dot}), these can be rewritten as
\begin{eqnarray}
\dot{g}_{rr} & = &  2 g_{rr} \beta ' - 2 N K_{rr} -8 \frac{g_{rr}f_{rT}}{g_T}\beta + 2 f_{rrr}\beta
\label{grrdot_b} \ ,\\
\dot{g}_T & = & -2 N K_T + 2\beta f_{rT} \label{gTdot_b}
\, . \end{eqnarray}
Note that since $g_{rr}$ and $g_{T}$ correspond to ingoing modes, we are free to choose
them such  that they satisfy Eqs.~(\ref{grrdot_b},\ref{gTdot_b}). As was mentioned,
one does not need to solve these two differential equations at the
boundary, since time derivatives are all that is required for the time update in the
method of lines, which we use for the time update (see section IV). Thus, we
simply use the expressions (\ref{grrdot_b},\ref{gTdot_b}) as the right hand side of the
corresponding field variable at the boundary points (and similarly for $\dot{f}_{rT}$
and $\dot{K}_{T}$, see Eqs.~(\ref{frTdot_b}, \ref{KTdot_b}) below).

Lastly, $C_2=0$ needs to be enforced at the boundary. Using its definition, we have
$$
\frac{1}{\sqrt{g_{rr}}}f_{rT}' - K_T' + l.o. = 0  \; .
$$
Trading spatial for time derivatives, this in turn gives
\begin{equation}
\frac{1}{\sqrt{g_{rr}}}\dot{f}_{rT} - \dot{K}_T + l.o. = 0     \; .
\label{c2_boundary}
\end{equation}

Next observe that, by definition, $\dot{K}_T=\dot{K}_T(\dot{u}_4,\dot{u}_6,
\dot{g}_{rr})$ and $\dot{f}_{rT}=\dot{f}_{rT}(\dot{u}_4,\dot{u}_6,
\dot{g}_{rr})$. Since $\dot{g}_{rr}$ at the boundary
is readily known from equation (\ref{grrdot_b}), Eq. (\ref{c2_boundary}) fixes the
 ingoing mode $\dot{u}_4$ (up to now free). From this,  the definition of $u_4$ and $u_6$,
and Eq. (\ref{grrdot_b}), we end with (recall that $u_6$ is outgoing, so it does not need
boundary conditions)

\begin{eqnarray}
& & \dot{f}_{rT} =    \frac{g_{rr}^{1/2}}{2} \dot{u}_6 + l.o.
\label{frTdot_b} \\
& & \dot{K}_{T} =  \frac{1}{2} \dot{u}_6 + l.o.
\label{KTdot_b}
\end{eqnarray}

Similarly, the ingoing mode $u_3$ is completely arbitrary. From it and the outgoing mode
$u_5$ we have
\begin{eqnarray}
K_{rr} & = & \frac{1}{2}(u_3+u_5)  \label{Krr_b} \; , \\
f_{rrr} & = &\frac{g_{rr}^{1/2}}{2}(u_5-u_3) \label{frrr_b} \; .
\end{eqnarray}

Finally, the ingoing mode $u_8$ is also arbitrary. From it and the outgoing mode
$u_7$ we have
\begin{eqnarray}
\Pi &=&  \frac{1}{2}(u_7+u_8)  \label{Pi_b} \; ,\\
\Phi &=& \frac{g_{rr}^{1/2}}{2}(u_7-u_8) \label{Phi_b} \; .
\end{eqnarray}

Equations
({\ref{grrdot_b},\ref{gTdot_b},\ref{frTdot_b},\ref{KTdot_b},\ref{Krr_b},\ref{frrr_b},\ref{Pi_b},\ref{Phi_b}})  completely determine the
boundary conditions for our eight variables; $u_3$ and
$u_8$ are the only free modes. In the vacuum case, $u_8=0$ and $u_3$ describes a gauge
mode.

In the next section we discuss our numerical implementation and results. One comment is
in order before that: there is an additional constraint, $C_{matter}=\Phi - \Psi '$, but
by repeating the treatment described above one can see that this constraint fixes the
boundary condition for $\Psi $: Since we are not evolving $\Psi $, we do not need to care
about this.

\section{Numerical simulations}

To implement the proposed boundary treatment strategy, we have chosen a straightforward
 second-order dissipative method of lines. In particular,
it was not necessary to employ techniques such as causal
differencing~\cite{causaldiff}, upwind discretization or any special way of treating
the Lie-derived terms in Eqs.~(\ref{grr_dot}-\ref{phidot}) (however, these can be readily
incorporated as well). Spatial derivatives
are discretized with second-order centered differences plus fourth-order dissipation as
discussed in~\cite{kreissoliger}, while for the time integrator we
use second-order Runge-Kutta (the dissipation added is, indeed, needed; otherwise,
this method is unstable even for a simple scalar wave equation; see, e.g., \cite{kreiss}).

Our uniform grid structure consists of points $i=0\ldots N$, with grid spacing
$\Delta r = L/N$, where $L= r_{outer}-r_{inner}$, and we implement derivatives according to standard
formulae
$$
Af'
\rightarrow AD_0 f - \frac{\sigma_i }{\Delta t} (\Delta r)^4 D_+D_+D_-D_-f  \; , $$
\begin{eqnarray*}
& & (D _0 f)_i =
\frac{f_{i+1} - f_{i-1} }{2\Delta r} \; , \\
& &  (D _+ f)_i =
\frac{f_{i+1} - f_{i} }{\Delta r} \; , \\
& & (D _- f)_i =
\frac{f_{i} - f_{i-1} }{\Delta r} \; .
\end{eqnarray*}

To evaluate derivatives at the boundaries, one either resorts to {\it one-sided}
derivatives (which require different algorithms applied at boundary points) or introduces
{\it ghost zones},  which are artificial points beyond the boundaries where field values
are defined via extrapolation (one can choose the
extrapolation order such that the answers from the different approaches are exactly
the same). For convenience we chose the latter approach with only one ghost-zone for each
boundary. Field values at these ghost zones are defined
via third-order extrapolations and the {\it same} derivative operator is applied $\forall
i, i=0..N$.

Since our inner boundary is always inside the black hole, we extrapolate all
variables at the ghost zone point ($i=-1$).

At the outer boundary ghost zone ($i=N+1$), the outgoing modes $u_5$ and
$u_7$ are also found by extrapolation, while the ingoing modes $u_3$ and $u_8$ are set as
arbitrary functions of time. Next, equations
({\ref{grrdot_b},\ref{gTdot_b},\ref{frTdot_b},\ref{KTdot_b}) are integrated, at each time
step, with second-order Runge Kutta. This, plus equations
(\ref{Krr_b},\ref{frrr_b},\ref{Pi_b},\ref{Phi_b}), completes the treatment for the outer
boundary.

As a side note, it is worth mentioning that since we use only one ghost zone for each
boundary, fourth-order derivatives cannot be obtained at the points $i=0,N$; thus, no
dissipation is added to these points (we set $\sigma_{i=0}=\sigma_{i=N}=0$).

In all the cases  discussed in this paper we have checked second-order
self-convergence for the eight main variables, for the constraints and mass, as well as
convergence with respect to the analytic solution, whenever this is available.
The Courant-Friedrich-Levy factor $\lambda = \Delta t/\Delta r$ is set to $0.25$, and the
dissipation factor is $\sigma_i = 0.6/16$ (for $i=1..N-1$).

\subsection{Vacuum evolutions}

\subsubsection{Stationary slicing of a black hole}

The simplest test of a black hole spacetime consists of reproducing known stationary
slicings of a Schwarzschild black hole. For this test we give the corresponding known
values both to the initial data and to the $u_3$ mode ($u_8$=0 for vacuum) at the outer
boundary. We concentrate here on Painlev\'e-Gullstrand
slicings, but we have obtained similar results using Kerr-Schild
slicings. The code runs for unlimited times, even with resolutions as coarse as $\Delta r
= M/6$ (with $M$ the mass of the black hole). Figure 2 displays the $L_2$ norm of the
Hamiltonian constraint for three different resolutions and $L=9M$ ($r_{inner}=M$), the
$L_2$ norm of a grid function $f$ being defined as $$
|f|:= \left( \frac{1}{L} \sum_{i=0}^{N-1}(f_i)^2\Delta r \right)^{1/2} \; .
$$
There
is no growth in the Hamiltonian constraint, and the same holds true for the other three
constraints.

\begin{figure}[ht]
\begin{center}
\epsfig{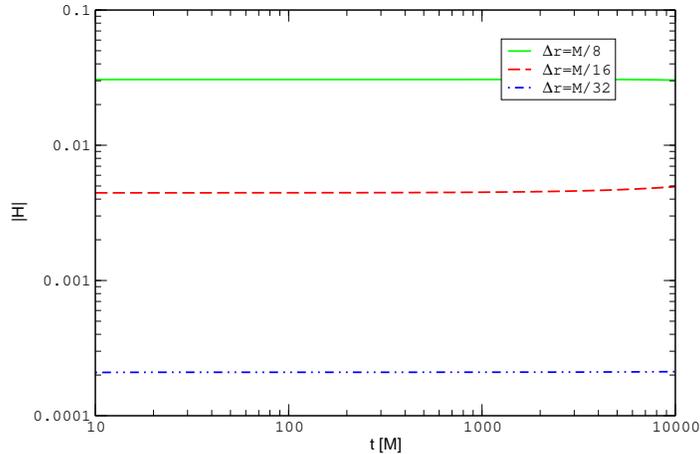}
\caption{$L_2$ norm of the Hamiltonian constraint, for the evolution of a
Painlev\'e-Gullstrand black hole. In these runs the outer boundary is at $r=10M$.}
\end{center}
\end{figure}

\begin{figure}[ht]
\begin{center}
\epsfig{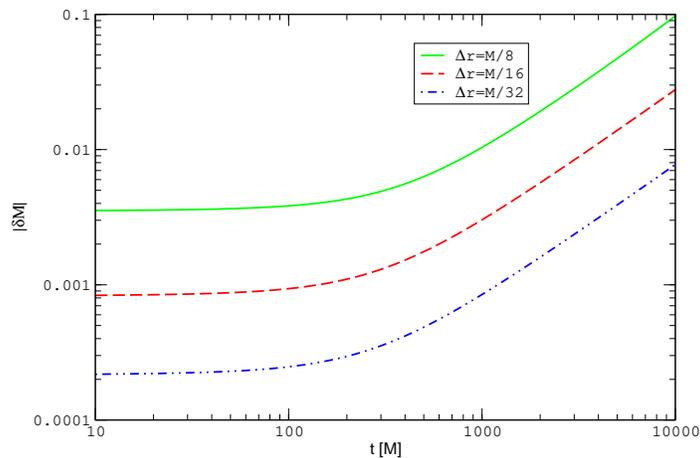}
\caption{$L_2$ norm of the relative mass error $|\delta M| = |{\cal M}-M|/M$, for the same
runs shown in figure 2. After one thousand crossing times, the relative error for the
coarsest resolution is around ten percent.}
\end{center}
\end{figure}

Figure 3 shows, for the same run, the $L_2$ norm of the relative error,
$|\delta M|$, in the mass function ${\cal M}$, defined as
$$
|\delta M|:= \frac{1}{M}|{\cal M}-M| \; ,
$$
where ${\cal M}$, the Misner-Sharp mass \cite{ms}, is a gauge invariant \cite{st}
definition of the ADM mass $M$ in spherical symmetric vacuum:

\begin{equation}
{\cal M} := \frac{rg_T^{1/2}}{2}\left[1+\frac{r^2}{g_T}\left(K_T^2-\frac{f_{rT}^2}{g_{rr}}
\right )\right]  \; .      \label{MS_mass}
\end{equation}
That is, each of the terms in the r.h.s. of Eq. (\ref{MS_mass}) might be a function of $t$
and $r$, but Einstein's vacuum equations in spherical symmetry imply that, at the
continuum, the r.h.s.is a constant (equal to $M$), both as a function of space and
time. When we compute $\delta M$ we evaluate ${\cal M}$ using the numerical values of the
right hand side of Eq. (\ref{MS_mass}), while for $M$ we use its analytical value.

From figure 3 we see that, as opposed to
the constraints, there is a drift in the mass (this effect is expected, as second-order
errors accumulate after many iterations) which, after a couple of hundred masses, is
essentially linear in time.
\begin{figure}[ht]
\begin{center}
\epsfig{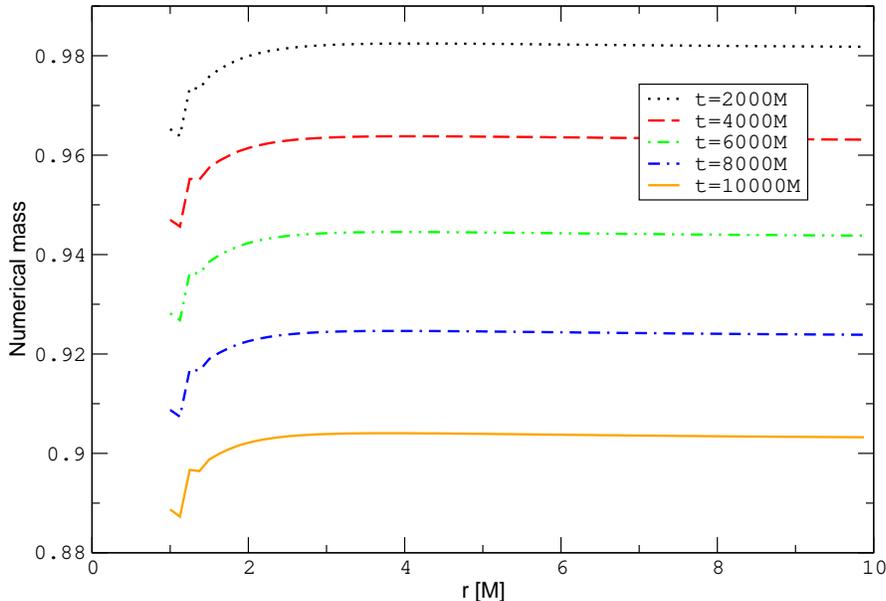}
\caption{Numerical value of the Misner-Sharp mass, scaled by its analytical value, as a
function of radius, for different times. The monotonic behavior is explained by the
fact that the black hole ``moves'' through the grid.}
\end{center} \end{figure}
In figure 4
we plot ${\cal M}/M$ as a function of radius, for different times and $\Delta
r=M/8$; the error is spread all over the domain and the mass decreases monotonically in
time, which  suggests that the black hole is moving through the grid (in the sense that
$r\rightarrow r+t\times v_{num}$). We have verified that this is indeed the case
(but, still, $v_{num} = {\cal O}\left(\Delta r ^2 \right)$) by computing the (grid)
location of the apparent horizon. For example, with $\Delta r=M/8$, after $10,000M$ the
horizon has moved from $r_{h}=2M$ to $r_h=1.75M$, which accounts for the roughly ten
percent error that the mass has at that time in figure 3: $$
\delta M \approx 1-r_h/(2M) \approx 0.1
$$

We have tried different positions for both boundaries, with the inner one always
inside the black hole and the outer one ranging from $10M$ to $1,000M$, and neither the
stability nor the convergence of the simulations depended on such positions. In figures 5
and 6 we show the analogue of figures 2 and 3; with the outer boundary placed at $50M$,
while still running the code up to $10^3$ crossing times (i.e.~$t=50,000M$).

\begin{figure}[ht]
\begin{center}
\epsfig{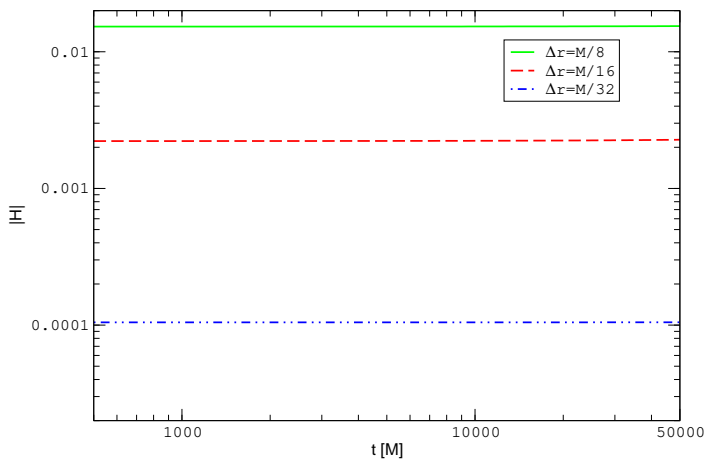}
\caption{Same as for figure 2, but now with the outer boundary at $r=50M$.}
\end{center}
\end{figure}

\begin{figure}[ht]
\begin{center}
\epsfig{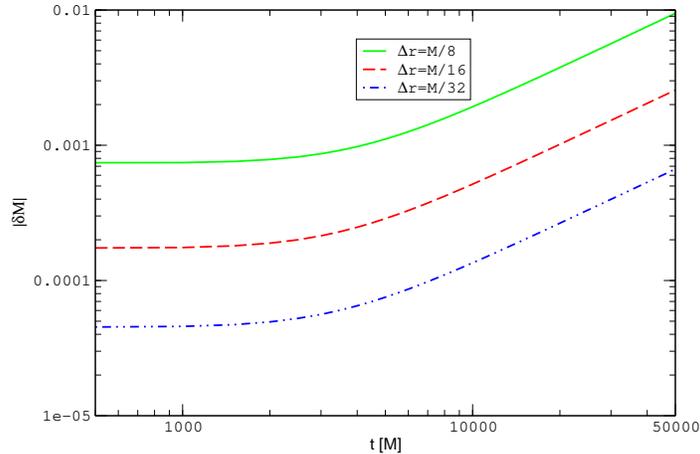}
\caption{Same as for figure 3, but now with the outer boundary at $r=50M$, as in the
previous plot.} \end{center}
\end{figure}

\subsubsection{Gauge pulse falling into black hole}
As discussed in Section III, the mode $u_3$ at the outer boundary is completely arbitrary.
In this next test we let a gauge pulse enter the domain through the boundary. We thus
use initial data corresponding to a Painlev\'e-Gullstrand slicing of a black hole of mass
$M$, and fix $u_3$ at $r_{outer}$ by superposing, on top of the Painlev\'e-Gullstrand
value, a Gaussian pulse described by
\begin{equation}
u_3(t) = u_3^{PG}\left(1+
Ae^{-(t-t_0)^2/\tilde{\sigma }^2} \right) \label{gauge_pulse} \, .
\end{equation}
\begin{figure}[ht]
\begin{center}
\epsfig{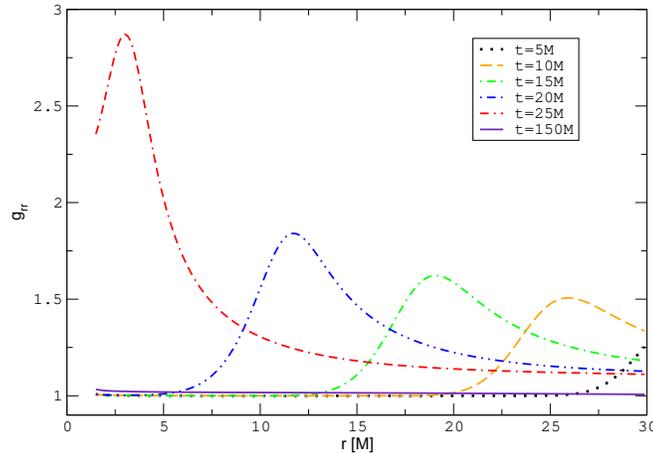}
\caption{The radial component of the metric, for a gauge pulse entering the domain
through the outer boundary. The amplitude's pulse grows as it approaches the inner
boundary, falls into the black hole, and finally the metric settles down to a stationary
one.}
\end{center}
\end{figure}

In figure 7 we present snapshots of the $g_{rr}$ component of the metric as a function of
radius for different times. The width of the pulse is $\tilde{\sigma } =2M$, and it is
centered at $t_0=5M$. The amplitude is quite large, $A=1$, corresponding to a $100\%$
``disturbance''  of the stationary value. The resolution employed is $\Delta r = M/8$, and the
domain extends from $r_{inner}=M$ to $r_{outer} =30M$. The pulse grows as
it approaches the inner boundary (which is not surprising, since it happens
even for a linear scalar equation propagating in a Schwarzschild background due to
the effective non-trivial potential), $g_{rr}$ having a
disturbance of more than $500 \%$ that of the stationary value when it
``crosses'' the inner boundary. After the pulse falls into the black hole, the metric
components gradually settle down to stationary values and the code runs for unlimited
times.

Since we are dealing
with a vacuum spherically symmetric spacetime, the resulting spacetime must be a
(dynamical) slicing of Schwarzschild. One way of corroborating this is to follow the
Misner-Sharp mass. As can be seen in figure 8, where $|\delta M|$ is shown as a function
of time for different resolutions, the numerical value of the mass does converge to this
analytical prediction. Just as in the stationary case, the code runs for as long as
wanted, and has a similar linear drift in the mass at late times. The growth in the errors
around $t=30M$ seen in figure 8 are not due to the pulse entering the domain through the
boundary (this happens at $t=5M$) but due to
the pulse reaching the inner boundary, where all the gradients are steeper, and
the huge growth in $g_{rr}$ is rather poorly resolved. Similar growths appear in the
constraints (see figure 9), but they are still second-order convergent. It is indicative
of the power of consistent boundary conditions that we can have such a big pulse entering
the domain through the boundary, causing the spacetime to be so dynamical, that the code
is stable and that the mass error is small (notice that even for a resolution as coarse as
$\Delta r=M/8$ this error, while the pulse travels trough the domain, does not exceed one
percent).

\begin{figure}[ht]
\begin{center}
\epsfig{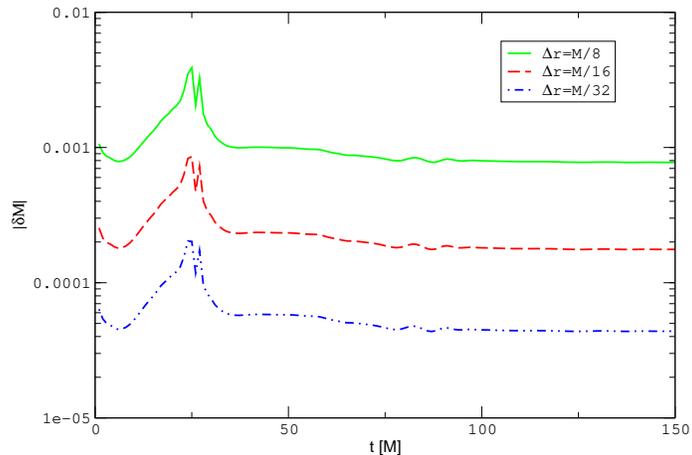}
\caption{Mass error for a gauge pulse ``falling'' into the black hole. The peaks around
$t=30M$ are caused by the pulse reaching the inner boundary. After the pulse falls
into the black hole the spacetime settles to a stationary one, and at late times the mass
has a linear drift, just as in the stationary evolutions. }
\end{center} \end{figure}

\begin{figure}[ht]
\begin{center}
\epsfig{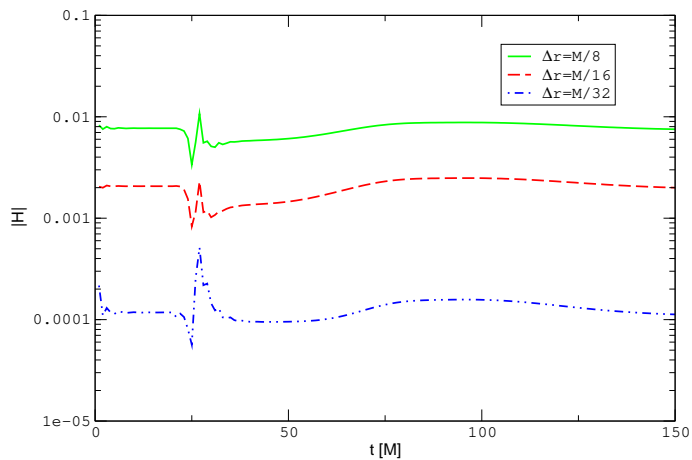}
\caption{$L_2$ norm of the Hamiltonian constraint for the same run shown in figure 8.}
\end{center} \end{figure}

In equation (\ref{gauge_pulse}) we wrote the boundary condition in that way (i.e.~the
Painlev\'e-Gullstrand value times a time dependent function) in order to have a measure of
how large the pulse is compared to its stationary value. In the next simulations we will
have a more physical idea of how strong these pulses can be in our code.

\subsection{Scalar field coupled to the black hole}
As a last test of the approach we consider the case of a minimally coupled
scalar field which also enters the computational domain through the outer boundary.  As
mentioned, only one scalar field mode is freely
specifiable at the outer boundary. It is this mode which is chosen to describe an incoming
pulse with amplitude $A$ and compact support in time as
$$
u_7(t) = A (t-t_I)^4 (t-t_F)^4 \sin(\pi
t) \, ,
$$
if $t\in[t_I,t_F]$, and $u_7(t) = 0$ otherwise. We choose
$u_3=0$, and for the initial data we take that of a Painlev\'e-Gullstrand black hole of
mass $M$. Note that one advantage of this is that one can
perform non trivial simulations without solving the constraints initially, since one can
provide, as we have done, a known solution of the constraints as initial data, and
introduce the non trivial dynamics through the boundary (where all the treatment is
algebraic), see also \cite{leco}.

We have tried different amplitudes and time
range for the pulse, obtaining stable discretizations in all cases. In order to illustrate
the robustness of the approach we here show two examples. In the first one we show
how the method can cope with considerably strong pulses (measured by the amount of energy
that the black hole accretes), and in the second one the tails of the scalar field
are computed and shown to agree with the expected result.

\begin{figure}[ht]
\begin{center}
\epsfig{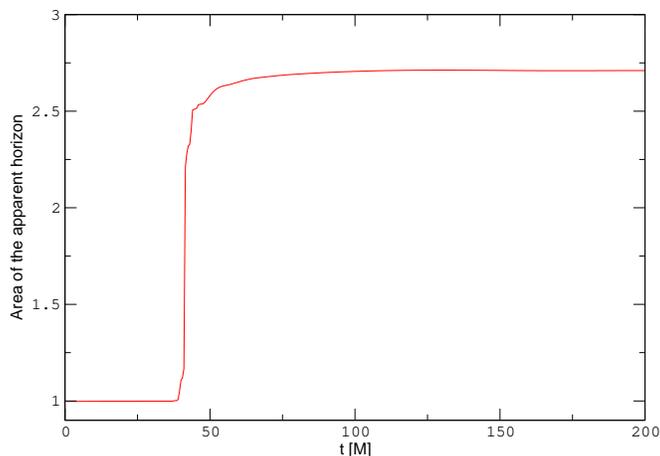}
\caption{Change in the area of the apparent horizon, due to a quite narrow ($\Delta t=
10M$) pulse of scalar field injected at the outer boundary, with a not very demanding
resolution ($\Delta r=M/8$). The final mass of the black hole is more than twice the
original one.}
\end{center}
\end{figure}

For the first case we chose $A=8$, $t_I=0, t_F=10M$, $\Delta r =
M/10$ and the outer boundary at $r=50M$. Figure 10 illustrates the area of the apparent
horizon as a function of time. Initially the area is $M$ and, as time progresses, it
increases by $270\%$, until it reaches a stationary state describing, as expected, a
Schwarzschild black hole of larger mass.

In the second case, in order to resolve the tails accurately, considerably finer resolution is needed.
For this reason, we chose $\Delta r = M/50$ and place the outer boundary at $r=100M$. In order to
measure the tails, we place four different observers: at $r=50, 70, 90$ and $100M$. The
  decay rate (of the form $\Phi \propto t^{-n}$) found by these observers is
$n=-2.89,-2.97,-2.94$ and $-3.04$, respectively. These are in excellent agreement with the
expected value of $n=-3$~\cite{tails}. Furthermore, the clean treatment of the outer
boundary allows for accurate measurements even at the last point of the computational
domain. Past works, which resorted to approximate boundary conditions, have observed
that the boundary influenced the results when the observers were placed close to it.

\begin{figure}[ht]
\begin{center}
\epsfig{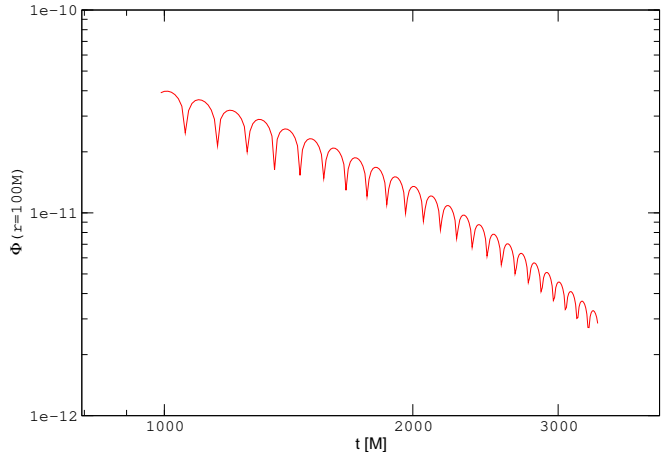}
\caption{Tail decay for the (time derivative of the) scalar field. The observer is
located at the last grid point ($r=100M$) and measures a tail decay of $\Phi \propto
t^{-3.04}$, in good agreement with the expected decay of $t^{-3}$. This is so
mainly because of the clean treatment for the boundary.} \end{center} \end{figure}

Finally, we should mention that we have tried with different kinds of time dependent
boundary conditions for the modes $u_3$ and $u_7$, and the results presented here do not
depend on the particular choice made. Not only can we define pulses that last for a finite
amount of time, as we have shown here, but we can define a time dependency that lasts
forever (say, $u_3= A\sin( \omega t)$). In this sense, the large growth in the mass of the black
hole mentioned above is what one can do with a pulse of width
$10M$. In fact, one can increase arbitrarily the mass by using wider
pulses, i.e.~by slowly injecting scalar field energy through the boundary.

\section{Conclusions}
In the present work we have presented additional support for the need to consider the
constraint equations when defining boundary conditions. Although approximate boundary
conditions (which are simpler to implement) can certainly be arranged to produce reliable
simulations, for long term evolutions they have some important disadvantages:
\begin{itemize}
\item{The inherent error introduced might accumulate,
spoiling the physical outcome: For instance, even in the particular case of a binary black
hole collision, the radiation is expected to be around $~5\%$ of the total mass of the
system. Therefore, even small errors which accumulate over time can account for a
significant percentage of the radiated energy. Since the outcome of some of these
simulations will be used as templates for gravitational wave data
analysis\cite{hughesflanagan}, one should minimize all possible waveform contaminating
sources and diminish both amplitude and phase errors.}
\item{Although the inherent error of approximate boundary conditions can be
reduced by placing the outer boundary farther away from the sources, this
amounts to having to compute the solution on a larger computational domain which 
naturally increases the cost of the simulation.}
\item{Standard approximate boundary approaches pay little or
no attention to the satisfaction of the constraints at the boundary. It is hard to
conceive that they do not introduce constraint violating modes into the solution. Unless
the implemented scheme is capable of damping these modes away, they will be present in
the computational domain. A possible source of instabilities (though not necessarily the
only one) in present implementations are, precisely, constraint violating modes, hence, it
is important to eliminate all possible  sources for these modes. It might happen that, in
certain cases, the observed constraint violations are due to those particular
implementations. Nevertheless, as the goal is to solve Einstein equations,  sources of
constraint violations must be removed. As mentioned, standard boundary treatments  are
likely to introduce these violations.}
\end{itemize}
Furthermore, carefully designed boundary conditions are important in the case of {\it artificial
boundaries}. These are boundaries which occur inside the computational domain
where the computational mesh has regions with different resolutions (for instance in the case of adaptive
mesh refinement) or  has been subdivided into sub-domains which are treated
independently (e.g. when using domain decomposition techniques). As these boundaries are
purely artificial and introduced for convenience, it is imperative to count with a clean
boundary treatment which eliminates (or at least minimizes) any spurious reflections or
numerical noise.

Consequently, the search for accurate boundary conditions is of importance in present applications.
The results presented in
this work attest to that effect and suggest a practical and simple way to obtain such
boundary conditions, which ensures constraint violating modes are absent by the very
definition of the approach. Current work is devoted to applying the same techniques to 3D
problems.

\section{Acknowledgments}

This work was supported by grants
NSF-PHY-0090091, NSF-PHY-9800973, by Fundaci\'on Antorchas, and by the Eberly Family
Research Fund at Penn State. It is a pleasure to thank O. Reula for discussions and
helpful insights, and M. Choptuik, B. Kelly, L. Kidder, J. Pullin, R. Matzner, P. Laguna,
M. Scheel, H. Shinkai, and S. Teukolsky for comments and suggestions. G.C. acknowledges E.
Frymoyer for financial support, and L.L. acknowledges financial support from CITA (as a
CITA National Fellow) and wishes to
thank the University of South Africa (UNISA) for its hospitality where parts of this work
were completed. Some computations were performed on the vn.physics.ubc.ca Beowulf cluster
which was funded by the Canadian Foundation for Innovation (CFI). The Center for Gravitational
Wave Physics is supported by the National Science Foundation under co-operative agreement
PHY 01-14375.

\appendix
\section{Evolution equations}

The main variables propagate according to
\begin{eqnarray*}
\dot{g}_{rr} & = & \beta g_{rr}' + 2g_{rr}\beta ' - 2\tilde{\alpha }g_{rr}^{1/2}g_TK_{rr}
\\
& & \\
\dot{g}_{T} & = & \beta g_T' - 2\tilde{\alpha }g_{rr}^{1/2}g_TK_T + \frac{2\beta g_T}{r}
\\
& & \\
\dot{K}_{rr} & = &  \beta K_{rr}'-\tilde{\alpha }g_{rr}^{-1/2}g_{T}f_{rrr}'- \tilde{\alpha }^{''}g_{rr}^{1/2}g_{T} -
6g_{T}^{-1}g_{rr}^{1/2}\tilde{\alpha }f_{rT}^2 + 4g_{T}r^{-1}g_{rr}^{1/2}\tilde{\alpha }'
- 6g_{T}r^{-2}g_{rr}^{1/2}\tilde{\alpha } + 2K_{rr}\beta ' -  \\
& & \\
& &
2g_{T}g_{rr}^{1/2}\tilde{\alpha }\Phi ^2 -
g_{T}g_{rr}^{-1/2}\tilde{\alpha }K_{rr}^2  +
2g_{rr}^{1/2}\tilde{\alpha}K_{rr}K_{T} -
8g_{rr}^{-1/2}\tilde{\alpha }f_{rT}f_{rrr} +
2g_{T}g_{rr}^{-3/2} \tilde{ \alpha }f_{rrr}^2
 +     \\
& & \\
& &
2g_{T}r^{-1} g_{rr}^{-1/2}\tilde{\alpha }f_{rrr} -
g_{T}g_{rr}^{-1/2}\tilde{\alpha }' f_{rrr}  \\
& & \\
\dot{K}_{T} & = & \beta K_T'  -
\tilde{\alpha }g_Tg_{rr}^{-1/2}f_{rT}' + 2\beta r^{-1}K_T + g_Tr^{-2}g_{rr}^{1/2}\tilde{\alpha } +
\tilde{\alpha }g_TK_TK_{rr}g_{rr}^{-1/2} - g_Tf_{rT}\tilde{\alpha }'g_{rr}^{-1/2} -
2\tilde{\alpha }f_{rT}^2g_{rr}^{-1/2}  \\
& & \\
\dot{f}_{rrr} & = &  \beta f_{rrr}' -\tilde{\alpha }g_{rr}^{1/2}g_{T}K_{rr}'
-4g_{rr}^{3/2}\tilde{\alpha }'K_T +
12g_{T}^{-1}g_{rr}^{3/2}\tilde{\alpha }K_Tf_{rT} -
4 g_{rr}^{1/2}\tilde{\alpha}K_Tf_{rrr} -
g_Tg_{rr}^{-1/2}\tilde{\alpha }K_{rr}f_{rrr} -  \\
& & \\
& &
10g_{rr}^{1/2}\tilde{\alpha }K_{rr}f_{rT} + 3f_{rrr}\beta '  +
g_{rr}\beta ^{''} - \tilde{\alpha} 'g_{rr}^{1/2}g_{T}K_{rr} +
2r^{-1}g_{T}g_{rr}^{1/ 2}\tilde{\alpha}K_{rr} +
8r^{-1}g_{rr}^{3/2}\tilde{\alpha}K_T +
4\tilde{\alpha }g_{rr}^{3/2}g_{T}\Phi \Pi \\
& & \\
\dot{f}_{rT} & = &  \beta f_{rT}' - \tilde{\alpha }g_{rr}^{1/2}g_TK_T' + \beta 'f_{rT} -
\tilde{\alpha }' g_{rr}^{1/2}g_TK_T + 2g_{rr}^{1/2}\tilde{\alpha }K_Tf_{rT} -
\tilde{\alpha }g_{rr}^{-1/2}K_Tf_{rrr}g_T +
2r^{-1}\beta f_{rT} \\
& & \\
\dot{\Phi } & = & \beta \Phi ' - \tilde{\alpha }g_{rr}^{1/2}g_T\Pi ' -
g_{rr}^{-1/2}\tilde{\alpha }g_T\Pi f_{rrr} +
2\tilde{\alpha }g_{rr}^{1/2}\Pi f_{rT} + 2r^{-1}\tilde{\alpha }g_{rr}^{1/2}g_T\Pi -
\tilde{\alpha }' g_{rr}^{1/2}g_T\Pi + \Phi \beta '  \\
& & \\
\dot{\Pi } & = & \beta \Pi ' - g_{rr}^{-1/2}\tilde{\alpha }g_T\Phi ' +
g_{rr}^{-1/2}\tilde{\alpha }g_T\Pi K_{rr} + 2\tilde{\alpha }g_{rr}^{1/2}\Pi K_T  -
4g_{rr}^{-1/2}\tilde{\alpha }\Phi f_{rT} + 2r^{-1}g_{rr}^{-1/2}\tilde{\alpha }g_T\Phi -
g_{rr}^{-1/2}g_T\Phi \tilde{\alpha }'
\end{eqnarray*}

\section{Excision of the singularity}

The issue of excision is a non trivial one, even if dealing with a strongly
hyperbolic  formulation of Einstein's equations, and even ignoring the constraints. The
point is that, depending on the formulation, modes can leave the black hole. Even though
these modes cannot be physical, they would need boundary conditions in any case in order
to fix the solution. If one, for example, extrapolated all variables in such a situation,
one would be implicitly be giving boundary conditions that depend on the discretization
and that might not have a consistent limit as the grid spacing is decreased. Here we show
a simple example to illustrate this point.

We start with a  formulation widely used in
numerical relativity, the ADM one (more precisely, the equations arising from $R_{ab}=0$),
and we choose exact lapse and exact co-shift (that is, the covariant shift). We use the
same notation as in the body of the paper, except that now $N=N(t,r)$ and $\beta _r =\beta
_r(t,r)$ are prescribed as arbitrary functions of spacetime. In spherically symmetry, the
evolution equations for such a formulation are
\begin{eqnarray}
\dot{g}_{rr} &=& -\frac{\beta _rg_{rr}'}{g_{rr}} + 2\beta _r' - 2NK_{rr}
\\ 
\dot{g}_{T} &=& \frac{\
\beta _rg_{T}'}{g_{rr}} + 2\frac{\beta _rg_T}{rg_{rr}} -
2NK_{T} \label{gtdot}\\
\dot{K}_{rr} &=&
-\frac{1}{2g_{rr}^2g_{T}^2r}\left( -2g_{T}^2r\beta_rK_{rr}'g_{rr} -
4g_{T}^2rK_{rr} \beta _r'g_{rr} + 4g_{T}^2rK_{rr}\beta _rg_{rr}' + \right.\\
& &  2N
g_{rr}^2 g_{T}^{''}rg_{T} + 4Ng_{rr}^2g_{T}'g_{T} - Ng_{rr}
g_{rr}'rg_{T} g_{T}' - 2Ng_{rr} g_{rr}'g_{T}^2 - Ng_{rr}^2
(g_{T}')^2 r + \\
& & \left. 2 Ng_{rr} K_{rr}^2 g_{T}^2r - 4Ng_{rr}^2 K_{rr}
g_{T}rK_{T} + 2N^{''}g_{rr}^2g_{T}^2r - g_{rr}'N
'g_{rr}g_{T}^2r \right) \\
\dot{K}_T & = &
\frac{1}{4r^2g_{rr}^2} \left( 4 \beta _r r^2 g_{rr}K_{T}' + 8 \beta _r r g_{rr}
K_{T} - 2 N g_{T}^{''} r^2 g_{rr} - 8 Ng_{T}'
 r g_{rr} + Ng_{rr}' r^2 g_{T}' + 2 Ng_{rr}' r g_{T} + \right. \\
& &  \left. 4 N
g_{rr}^2 - 4 N g_{rr} g_{T} + 4 N K_T r^2 g_{rr} K_{rr} - 2 r^2
N' g_{rr} g_{T}' - 4 r N' g_{rr} g_{T} \right )
\end{eqnarray}
Note that in the previous equations, only $g_T$ appears with second derivatives.
This means that we can rewrite this system as a first-order one by
introducing just one new variable, $z:= g_T'$; i.e.~$\dot {u}=Au' +B$, where
$u=(g_{rr},g_T,K_{rr},K_T,z)^{\dagger }$.

As an evolution equation
for $z$ we make the simplest possible choice: we just take the spatial derivative of the
r.h.s. Eq. (\ref{gtdot}) (i.e.~we do not add the constraints to this new
equation).  Also, we replace everywhere $g_T'$ by $z$. The principal part is, then:
$$
A=\left(
\begin{array}{ccccc}
\displaystyle{-\frac{\beta _r}{g_{rr}} }& 0 & 0 & 0 & 0 \\
& & & & \\
0 & 0 & 0 & 0 & 0 \\
\displaystyle{\frac{-4rg_TK_{rr}\beta _r+2Ng_{rr}g_T + Nzrg_{rr} +
rg_{rr}g_TN'}{2rg_{rr}^2g_T} }&
0 & \displaystyle{\frac{\beta _r}{g_{rr}}} & 0  &
\displaystyle{-\frac{N}{g_T}} \\
& & & & \\
\displaystyle{\frac{N(rz+2g_T)}{4rg_{rr}^2}}
& 0 & 0 & \displaystyle{ \frac{\beta _r}{g_{rr}} }& \displaystyle{-\frac{N
}{2g_{rr}}} \\ & & & & \\
\displaystyle{-\frac{\beta _r (rz+2g_T)}{rg_{rr}^2} }& 0 & 0 & -2N&
\displaystyle{\frac{\beta _r}{g_{rr}}}
\end{array}
\right)
$$
This matrix has as eigenvalues and eigenvectors
\begin{eqnarray}
& & u_1 = \left[1,0, - \frac{g_{rr}N'-4\beta _r K_{rr}}{4\beta _rg_{rr}},0,
\frac{rz+2g_T}{2rg_{rr}}\right] \;\;\;, \left( v_1=-\frac{\beta _r}{g_{rr}} \right) \\
& & u_2 = \left[0,0,1,0,0 \right] \;\;\;, \left( v_2= \frac{\beta _r}{g_{rr}} \right) \\
& & u_3 =  \left[0,1,0,0,0 \right] \;\;\; \left( v_3=0 \right) \\
& & u_4 = \left[0,0, -\frac{g_{rr}^{1/2}}{g_{T}}, - \frac{1}{2g_{rr}^{1/2}} ,1 \right]
\;\;\;, \left( v_4 = \frac{\beta _r + Ng_{rr}^{1/2}}{g_{rr}} \right) \\
& & u_5 = \left[0,0, \frac{g_{rr}^{1/2}}{g_{T}},  \frac{1}{2g_{rr}^{1/2}},1\right]
\;\;\; ,\left( v_5 = \frac{\beta _r - Ng_{rr}^{1/2}}{g_{rr}} \right)
\end{eqnarray}
The determinant of the matrix that diagonalizes $A$ is proportional to
\begin{equation}
\frac{(-4K_{rr}\beta _r + g_{rr}N
')(rz+2g_T)^2}{64g_{rr}^{7/2}\beta _r r^2} \; , \label{deter}
\end{equation}
the system being strongly hyperbolic unless this determinant is zero. For
the usual slicings of Schwarzschild (Painlev\'e-Gullstrand, Kerr-Schild, time harmonic
and full harmonic slicings) expression (\ref{deter}) is different from zero,
so the system is, indeed, strongly hyperbolic, and the same will hold for
(perhaps slight) distortions of those spacetimes. Now, it is known that one has to give initial data and
boundary conditions to the incoming characteristic modes in order to fix the
solution in such a system. The point is that here there is
always a negative eigenvalue (whose corresponding eigenmode will thus propagate in the
direction of increasing $r$, in particular, leaving the black hole): if $\beta _r \neq 0$,
these eigenmode is either $u_1$ or $u_2$, while the mode is $u_5$ if $\beta _r=0$.

It is important to point out that whether modes leave or not the black hole strongly
depends on the particular formulation that one is dealing with. For example, using exactly
the same formulation (ADM in spherical symmetry) with other choices
of lapse and shift (such that they `lock' the area) also gives strongly hyperbolic
formulations when rewritten in first order form, but they {\em do not} have modes leaving
the domain (see \cite{kelly}). Also, if one uses exact-lapse and exact-shift (as opposed
to exact co-shift), the system is only weakly hyperbolic and, thus, the characteristic
modes are not complete but, in any case, they do not leave the black hole
(see, also, \cite{kelly}). What we want to emphasize here  is that superluminal
modes can appear quite naturally, even in standard formulations.


\end{document}